\documentstyle[11pt,twoside]{article}

\begin{document}
\thispagestyle{empty}
\begin{center}

{\bf Large Extra Dimensios}\footnote{Talk presented at the
German-Caucasus School and Workshop on Hadron Physics (2004, Tbilisi)}\\
\vspace{2cm}
Ghela G. Devidze\footnote{devidze@hepi.edu.ge},
Akaki G. Liparteliani\footnote{lipart@hepi.edu.ge}\\
{\em High Energy Physics Institute, Tbilisi State University,\\ Tbilisi,
Georgia}
\end{center}
\vspace{0.5cm}
\bigskip

The modern status of large extra dimensional (LED) approaches, highlights of
their manifestation and short prehistory are discussed in this paper.

\newpage
The modern models [1-5] with extra space-time dimensions could be built in
several ways. Among them the following major approaches are most remarkable:\\
1) ADD model of Arcani-Hammed, Dimopoulos and Dvali (1998) [1]. In this
approach all elementary particles except graviton are localized on the Brane,
while the graviton propagates in the whole Bulk. There are at least two extra
space dimensions in the above model.\\
2) RS model of Randal and Sundrum with warped 5-dimensional space-time and
nonfactorized geometry [2]. There are two such models with compactifized
and noncompactifized dimensions (RS1 and RS2 models [2]).\\
3) ACD models of Appelquist, Cheng and Dobrecu (so called Universal Extra
Dimensional Model), where all the particles move in the whole Bulk [3].

The idea of extra dimensions is as old as almost century age. It was
G. Nordstrom, who in 1912-1914 has formulated relativistic theory in
5 dimensions, which simultaneously described gravity and electromagnetism [6].
Of course Nordstrom unification looked rather formal. Nevertheless the
possibility of this 5dimensional unification hinted on deep relation between
these two fundamental interactions, known in that time. 

In 1915 Einstein created the General Theory of Relativity [7], where gravity
was considered as geometrical deformation of the space-time. In 1919
mathematician Th. Kaluza has shown that five dimensional relativistic gravity
(with Einstein-Hilbert action) manifests itself in the four dimensional
space-time as both electromagnetism and gravity [8]. 

In other words, Nordstrom formulated first five dimensional electrodynamics,
while Kaluza created first 5D gravity. These two were just first nave
attempts towards first unification of interactions (gravity and
electromagnetism) established that time.

In 1926 O. Klein rediscovered Kaluza's theory [9]. Moreover, there were also
numerous efforts to unify electromagnetism, gravity and quantum mechanics
in 5D by several authors. Later, the discovery of new types of interactions
(strong and weak ones) complicated above efforts. The use of one additional
dimension seemed to be unnatural and insufficient for these goals. 

One of the difficulties of multidimensional theories is the mechanism, due
to which extra dimensions are hidden. Thus during study of ordinary physical
phenomena the space-time looks like effectively four-dimensional. Until
recently mostly Kaluza-Klein type theories were considered. In these types
of theories extra dimensions are assumed as essentially compact and in
essence homogenous. Just compactness of extra dimensions provides effective
4D character of the space-time dimension at the distances above the
compactification scale (size of extra dimensions). At that the extra
dimensions must be of microscopic size. Following to widespread opinion,
the compactification scale should be of the Planck scale size (though
electroweak scale have been discussed in this role). On the other hand,
direct observation of extra dimensions at the Planck scale
($l_{Pl} \sim 10^{-33}cm, M_{Pl} \sim 10^{19}GeV$) seems to be hopeless.

However "Brane World" conception permitted to change situation on this
direction: we mean just the localization of ordinary matter (with the
exception maybe gravitons and other hypothetical particles which interact
very weakly with the matter) on the three dimensional manifold which is
cold the Brane [10]. The Brane is embedded into ambient higher dimensional
manifold (Bulk). Extra dimensions in the Brane World approach may have
large and even infinitely large size, leading to experimentally observable
effects.

Recent development of multidimensional models is encouraged mainly due to
superstring theories and their generalization M-theory, which is only
consistent quantum theory containing (at least in principle) all interactions
including gravity for today. Both superstring and M-theory most naturally are
formulated in the d=10 and d=11 dimensions correspondingly. Just this latter
circumstances  indicate the possibility of the existence of extra dimensions. \\

There are no experimental evidences in favor extra dimensions yet. From the
phenomenological  point of view the driving forces for modern extra
dimensional approaches are connected with the existence of hierarchy problem
  ($M_Z \ll M_{Pl}$) and with that of non vanishing cosmological
$\Lambda$  term ($\Lambda\sim 10^{-48}GeV^4$). It is very hard to explain
such a small but nonzero $\lambda$ value in the framework of 4D theory.
However, it would be mentioned that none of above phenomenological
motivations could be considered as a direct indication of the ultimate prove
in favor of extra spatial dimensions. For example: hierarchy
($M_Z \ll M_{Pl}$) has beautiful explanation in frame of 4D GUTs,  and
convincing solution of $\Lambda$-problem is not found in multidimensional
theories yet, though there are very interesting new approaches in this
direction.

Let us discuss the model, which illustrates the new understanding of
hierarchy problem [1]. Let us consider the question, how gravity for the
particles on the Brane becomes four dimensional. There are some answers on
this question. Simple possibility to answer the question is that extra
dimensions are compact and are characterized by size R. Gravity in this
model is four dimensional at $r \gg R$, but stops to be such at $r \sim R$.
One has N-dimensional Newton law at $r \ll R$: $V(r)=G_Nm_1m_2/r^{1+n}$
($G_N$ being fundamental gravitational constant in N+1 dimensional
space-time, n=N-3 being the number of extra dimensions). For $r \gg R$ the
four
dimensional Newton law works: $V(r)=Gm_1m_2/r$. Lacing potentials at
$r \sim R$ leads to the conditions $GR^n=G_N$. Introducing some fundamental
mass parameter M, which is connected with $G_N$ as $ G_N=1/M^{2+n}$, we
have $M_{Pl}=M(RM)^{n/2}$.

So, 4D gravity coupling G and $M_{Pl}$ are effective values and $M_{Pl}$
could be different from the new mass parameter M. This allows us to solve
the hierarchy problem in the new unexpected way. One can assume that
fundamental scale M coincides with electroweak scale by order of magnitude.
In this case lacing condition will define the size R of extra dimensions.
So, putting $M\sim 1TeV$, we have $r=10^{30/n-17}cm$. It is evident that
the case n=1 leads unacceptably large size of $R\sim 10^{13} cm$, while
already n=2 gives $R\sim 0,1 mm$. This sub millimeter region is just most
interesting for checking Newton law at small distances (the Newton law
behavior changes in the model when $r\sim R$ and sub millimeter region is
accessible for modern experiments). The case n=3 leads to $R\sim 10^{-7}$ cm
and for $n>3$ we have even smaller R. So, the deviations from Newton law at
distances of the order R becomes more difficult
to detect (if ever possible).

One should be stressed that such approach to the hierarchy problem rewrites
the problem in different language rather than solves it. Now the problem
sounds as follows: why the size of extra dimensions is large comparing with
new fundamental scale $l\sim M^{-1}\sim  10^{-17} cm$.

Remarkable feature of this new insight to the hierarchy problem is that
gravitational interactions become sizable not at the Planck scale, but on
the new scale of $M\sim O(TeV)$, which in fact must be considered as only
fundamental scale of the nature. Such alternative permits us to explore the
TeV scale region at the forthcoming accelerators, including from the point
of viability of extra dimension approaches.

If so, the LHC accelerator at CERN, as well as $e^+e^-$ colliders with
center of mass energies $\sim$ O(TeV) will be able to comb out thoroughly
this region of new fundamental scale of interactions (including gravity) up
to energy region of several TeV-s. Interesting accelerator effects of extra
dimensions are based on the virtual KK-graviton exchange phenomena. The
search for above processes on the new generation accelerator facilities
would be sensible to the gravity interaction scale M in the range of few
TeV-s. In fact light KK-gravitons are model independent feature of ADD
model. If really $M\sim O(TeV)$ is the fundamental scale of the theory, one
could expect very rich physics at above scale. Particularly, one can expect
existence of new particles with the masses of O(M), which could live on
Brane or move in the volume out of it.

ADD scenario predicts interesting phenomena in the few TeV range. In
particular, above mentioned feature of the possible presence of light KK
gravitons could manifest itself in the processes
$e^+e^\rightarrow \gamma + E$, $qq'\rightarrow > jet+E$ either indirectly
or through contact interactions caused with virtual KK-exchanges.
This two processes could be inspired just by virtual KK exchanges, providing
us with the information on the physics of extra dimensions. So, the LEP-II
single photon production has been measured already and for the case of n=2
one can find: $R<0.48mm$, $M>1200GeV$ at $95\%$ confidence level [11].
CDF collaborations has presented analogous bound basing on monojet
production analysis. Their limit for n=2 case is $R<1.2mm$ and $M>750GeV$
correspondingly [11].

Models of RS1 type lead to the exponential hierarchy between weak and Planck
scales (for comparison only fundamental scale in ADD approach actually is
the weak scales nearby region). Likewise to ADD approach gravity becomes
strong at the few TeV order energy range. However manifestations of new
features of gravity at accelerator experiments differ from each other in
the two approaches. The reason for this difference is situated mainly by
completely different graviton spectra in ADD and RS1 models. The
distinctive features of the RS1 phenomenology at TeV scales are
gravitational resonances which interact significantly with ordinary
particles.

Universal Extra Dimensions (UED) models could have very rich low energy
phenomenology [3]. Above compactification scale UED are higher dimensional
field theories whose equivalent description in 4D space-time includes
Standard Model (SM) KK-towers of SM fields and additional towers of KK-modes
which have not corresponding partners in the SM. All SM fields have their
heavy KK partners. The simplest UED model is the Appelquist, Cheng,
Dobrescu (ACD) model [3] with only extra dimension. Important feature of
ACD model is  the conservation so called KK-parity. This latter feature
excludes KK-contributions into FCNC processes at the tree level. On the
other hand the same FCNC inspired processes (like oscillations and rare
decays) on the first appear just at the one loop level in the SM. So, ACD
contributions have big chances to compete  and even exceed  corresponding
SM ones[3,9]. Indeed, calculations show that processes ,
$B_{s,d}\rightarrow \mu^+\mu^-$,  $B_{s,d}\rightarrow X_{s,d}\nu\bar{\nu}$ ,
$K_L\rightarrow \mu^+\mu^-$ , $B\rightarrow X_s \mu^+\nu^-$ ,
$K^+\rightarrow \pi^+\nu\bar{\nu}$, $K_L \rightarrow\pi^o \nu\bar{\nu}$
enhance considerably in the certain conditions [12].

The present limit (accelerator experiments) on the universal extra dimension
scenario is rather weak, as weak as $1/R>250GeV$ for one extra dimension
(ACD) from precision data. For case of two extra dimensions the limit
depends logarithmically on cut off scale. Roughly, the limit is around
400-800 GeV [11].

Theories with extra dimension seem a bit  exotic for today, especially
because they pretend to decrease beyond SM scales down to $\sim$few TeV.
LED approaches lead both to important conceptions about what can happen on
the scales accessible for future accelerators and to rare effects of low
energy physics. In the framework of these approaches plenty of
phenomenological questions could be considered. Moreover, there arise new
ideas on the road to solve the problems of fundamental importance, such as
($\Lambda$-problem or the problem of evolution of our Universe. All this
makes very interesting subject, shortly discussed here and stimulates
further experimental and theoretical research in this direction.

The research described in this publication was made possible in part by
Award No. GEP1-33-25-TB-02 of  the Georgian Research and Development
Foundation (GRDF) and the U.S. Civilian Research $\&$ Development Foundation
for the Independent States of the Former Soviet Union (CRDF). We thank
G. Bellettini, I.I. Bigi, G. Dvali, G. Gabadadze, C. Kolda, T. Lomtadze,
N.Nokolaev, G. Volkov for useful discussions and support. Discussions on
the German-Caucasus School and Workshop on Hadron Physics (2004, Tbilisi)
and IHEPI-seminar is acknowledged by authors also.\\

\section*{References}

1.~~ N. Arkani-Hamed, S. Dimopoulos, G. Dvali, Phys.Lett. B429 (1998) 263; hep-ph/9803315.\\
~~~~ I. Antoniadis, N. Arkani-Hamed, S. Dimopoulos, G. Dvali, Phys. Lett. B436 (1998) 257; hep-ph/9804398.\\
2.~~ L. Randall, R. Sundrum, Phys. Rev. Lett. 83 (1999) 3370; hep-ph/9905221.\\
~~~~L. Randall, R. Sundrum, Phys. Rev. Lett. 83 (1999) 46090; hep-th/9906064.\\
3. ~~T. Appelquist, H.C. Cheng, B.A. Dobrescu, Phys. Rev. D64 (2001) 035002; hep-ph/0012100.\\
4. M. Gogberashvili, Europhys. Lett. 49 (2000) 396; hep-ph/9812365.\\
5. G. Dvali, G. Gabadadze, M.Porrati, Phys.Lett. B485 (2000) 208; hep-th/0005016.\\
6. Nordstrom G, Phys. Ztsch., 1914, Bd. 15, S.504.\\
7. A. Einstein, Published in Sitzungsber. Preuss.Akad.Wiss.Berlin (Math.Phys.) 1915:778-786,1915, 
Addendum-ibid.1915:799-801,1915\\
8. Th. Kaluza.  Sitzungsber. Preuss.Akad.Wiss.Berlin(Math.Phys) 1921 (1921) 966.\\
9. O. Klein.   Z.Phys.37 (1926) 895.\\
10. V.A. Rubakov, M.E. Shaposhnikov, Phys. Lett. B125 (1983) 136.\\
~~~~K.Akama, in Gauge Theory and Gravitation: Proc. of the Intern. Symp., Nara, Japan, 1982 
(Lecture Notes in Physics, Vol. 176, Eds. K.~Kikkawa, N.~Nakanishi, H.~Hariai)(Berlin:
 Springer-Verlag, 1983), p.267.\\
~~~~I.~Antoniadis, Phys. Lett. B246 (1990) 377.\\
11. K. Cheung, hep-ph/0409028.\\
12. A.J. Buras, A. Poschenreider, M. Spranger and A. Weiler, hep-ph/0307202.\\

\end{document}